\documentclass[preprint2, 11pt]{aastex62}
\usepackage[utf8]{inputenc}
\usepackage{rotating}
\usepackage{natbib}
\usepackage{url}
\newcommand\mjup{M$_{Jup}$\ }
\newcommand\um{$\mu$m}
\usepackage{layout}

\begin{document}
\onecolumngrid
\begin{center}
\vspace{2in}
{\bf\Large A White Paper Submitted to \\
\vskip 0.15 in
The National Academy of Science's \\
\vskip 0.15 in
Committee on Exoplanet Science Strategy: \\
\vskip 0.15 in
Observing Exoplanets with the}\\
\vskip 0.15 in
{\bf\Large James Webb Space Telescope}\\
\vskip 0.15 in
{\bf\Large March 9, 2018}
\end{center}
%\thispagestyle{empty}
%\onecolumngrid 
\vskip 0.15 in
{\bf Charles A. Beichman$^{1,8,20}$, 
Tom P. Greene$^{2}$, }
David  Barrado y Navascu\'es$^{28}$, 
Natalie Batalha$^{2}$, 
Rus Belikov$^{2}$, 
Bjorn  Benneke$^{11}$, 
Zachory Berta $^{29}$, 
Elodie Choquet$^{8}$, 
David Ciardi$^{1}$, 
Ian Crossfield$^{9}$, 
Leen Decin$^{21}$, 
Jean-Michel  Desert $^{23}$, 
Rene Doyon$^{11}$, 
Courtney Dressing$^{25}$, 
Jonathan Fortney$^{14}$, 
Peter Gao$^{25}$, 
Andras Gaspar$^{12}$, 
Sasha Hinkley$^{27}$, 
Andrew Howard$^{20}$, 
Steve Howell$^{2}$, 
Renyu Hu$^{8}$, 
Nicolas Iro$^{18}$, 
Tiffany Kataria$^{8}$, 
Laura Kreidberg$^{7}$, 
David Lafreni\`ere$^{11}$, 
Pierre-Olivier Lagage$^{5}$, 
Fred Lahuis$^{24}$, 
Gregory Laughlin$^{26}$, 
Michael Line$^{4}$, 
Jack Lissauer$^{2}$, 
Mercedes Lopez-Morales$^{7}$, 
Jonathan Lunine$^{6}$, 
Luigi Mancini$^{19}$, 
Avi Mandell$^{16}$, 
Mark Marley$^{2}$, 
Michael Meyer$^{13}$, 
Cyrine  Nehm\'e$^{22}$, 
Rebecca Oppenheimer$^{3}$, 
Laurent Pueyo$^{10}$, 
Seth  Redfield$^{30}$, 
George Ricker$^{9}$, 
George Rieke$^{12}$, 
Marcia Rieke$^{12}$, 
Dimitar Sasselov$^{7}$, 
Everett Schlawin$^{12}$, 
Josh Schlieder$^{16}$, 
Sara Seager$^{9}$, 
Remi Soummer$^{10}$, 
Kevin Stevenson$^{10}$, 
Kate Su$^{12}$, 
Mark Swain$^{8}$, 
Kamen  Todorov$^{23}$, 
Olivia Venot$^{15}$, 
Gillian Wright$^{17}$, 
Marie Ygouf$^{1}$.

\vskip 0.15in
$^{1}$IPAC/NASA Exoplanet Science Institute;
$^{2}$NASA Ames Research Center;
$^{3}$American Museum of Natural History;
$^{4}$Arizona State University;
$^{5}$CEA Saclay;
$^{6}$Cornell University;
$^{7}$Harvard-Smithsonian Center for Astrophysics;
$^{8}$Jet Propulsion Laboratory, California Institute of Technology;
$^{9}$Massachusetts Institute of Technology;
$^{10}$Space Telescope Science Institute;
$^{11}$Universit\'e de Montr\'eal;
$^{12}$University of Arizona;
$^{13}$University of Michigan;
$^{14}$University of California, Santa Cruz;
$^{15}$LISA, CNES, France;
$^{16}$Goddard Spaceflight Center;
$^{17}$U.K. Astronomy Technology Centre,UK;
$^{18}$University of Vienna;
$^{19}$MPIA, Heidelberg;
$^{20}$California Institute of Technology;
$^{21}$University of Leuven, Belgium;
$^{22}$Notre Dame University, Lebanon;
$^{23}$University of Amsterdam ;
$^{24}$SRON, Netherlands;
$^{25}$University of California, Berkeley;
$^{26}$Yale University;
$^{27}$University of Exeter, UK;
$^{28}$Centro de Astrobiología,Spain;
$^{29}$University of Colorado, Boulder;
$^{30}$Wesleyan University.

%\end{document}
%\maketitle
%\clearpage

\clearpage\maketitle

\twocolumngrid 

\setcounter{page}{1}
 \section{Abstract}
 %vskip -0.15in
The James Webb Space Telescope (JWST) will revolutionize  our understanding of exoplanets with transit spectroscopy  of a wide range of mature planets close to their host stars  ($<$2 AU) and with coronagraphic imaging and spectroscopy of  young objects located further out ($>$10 AU). The census of exoplanets has revealed an enormous variety of planets orbiting stars of all ages and spectral types. With   TESS    adding to this  census with its all-sky   survey of  the closest, brightest stars,  the challenge of the coming decade will be to move from demography to physical characterization.  This white paper discusses the wide variety of exoplanet opportunities enabled by JWST's sensitivity and stability, its high angular resolution, and its suite of powerful instruments.  JWST observations will advance our understanding of the atmospheres of young to mature planets and will provide new insights into planet formation.

%end{document}

 %vskip -0.15in
\section{Introduction}
 %vskip -0.15in
The exoplanet community has made good first steps in characterizing the physical  and chemical  properties of a small number of planets, but many important questions remain including:
%with three different techniques: combining transit photometry with precision radial velocity (PRV) measurements (yielding planetary radius, mass, and density); measuring the brightness and/or spectrum of a planet during transit, secondary eclipse or through a full orbit for atmospheric characterization; and photometric or spectroscopic observations of directly imaged planets to study their bulk atmospheres.  

%But many important questions remain including:
\begin{itemize}
\item How does atmospheric composition vary as a function of key exoplanet characteristics, such as mass, radius, level of insolation and location within the planetary system?
\item What can we learn about vertical structure of exoplanet atmospheres, cloud formation and global atmospheric motions?
\item What can we learn about the formation of exoplanets from, for example, differences in their atmospheric carbon-to-oxygen (C/O) ratios  or overall metallicities  compared to those of their host stars? 
\item Do massive planets found on distant orbits ($>$50 AU) have a different formation mechanism compared to those on closer orbits?
\item Can we distinguish between low vs. high initial entropy states for forming massive planets to test two different formation possibilities?
\item What are the atmospheric constituents of mini-Neptunes, super-Earths, and even terrestrial planets?
\item How does the presence of planets affect the structure of debris disks?
%\item What can the composition of debris disk material tell us about the formation of planets?
\item Do young planetary systems with  known massive planets also contain lower mass planets?
\end{itemize}

JWST is poised to make dramatic advances in all of these areas. The ideas described here are drawn primarily from projects being proposed by the Guaranteed Time Observers (GTO) and the recently selected Early Release Science (ERS) observers. The broader  community will come up with many new projects, some of these before launch, but the most important and most innovative uses of JWST will doubtless come after launch as we learn about the telescope and the instruments and more importantly as we learn more about exoplanets themselves.
 
 %vskip -0.15in
\section{Direct Imaging and Spectroscopy}
 %vskip -0.15in
%\subsection{Planet Characterization}

{\bf 3.1 Planet Characterization:} JWST's great strength for direct imaging of planets comes from the stability of its Point Spread Function over a time scale of many hours and its operation in the low background space environment. The smallest Inner Working Angles (IWA) of the NIRCam coronagraphs are  relatively coarse, $4\lambda$/D  corresponding to 0.4\arcsec\ $\sim$  0.6\arcsec\ at 3.0 and 4.4 \um. The effective IWA for the MIRI Four Quadrant Phase Masks (4QPMs) is roughly $\lambda$/D, or 0.4\arcsec\ at 11.4 \um. Contrast limits at 1\arcsec\ are predicted to approach 10$^{-5}$  for NIRCam  and 10$^{-4}$ for MIRI (Beichman et al 2010). The on-orbit performance of the coronagraphs will depend sensitively on the drift in telescope wavefront error (WFE) over the duration of an observation as well as on the ability of the telescope to hold the star centered on the coronagraphic masks.

\begin{figure*}
\includegraphics[width=0.45\textwidth]{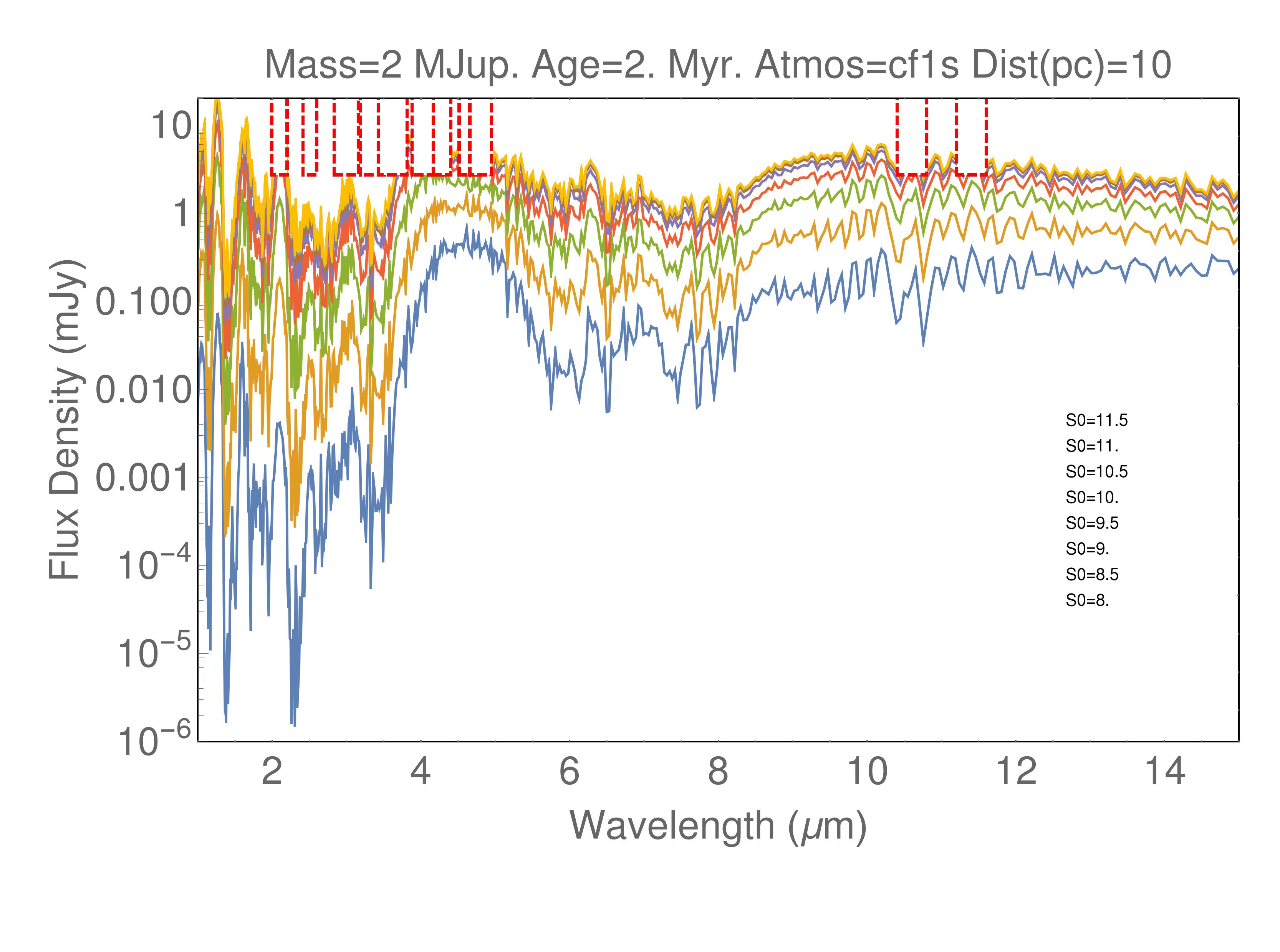}\includegraphics[width=0.4\textwidth]{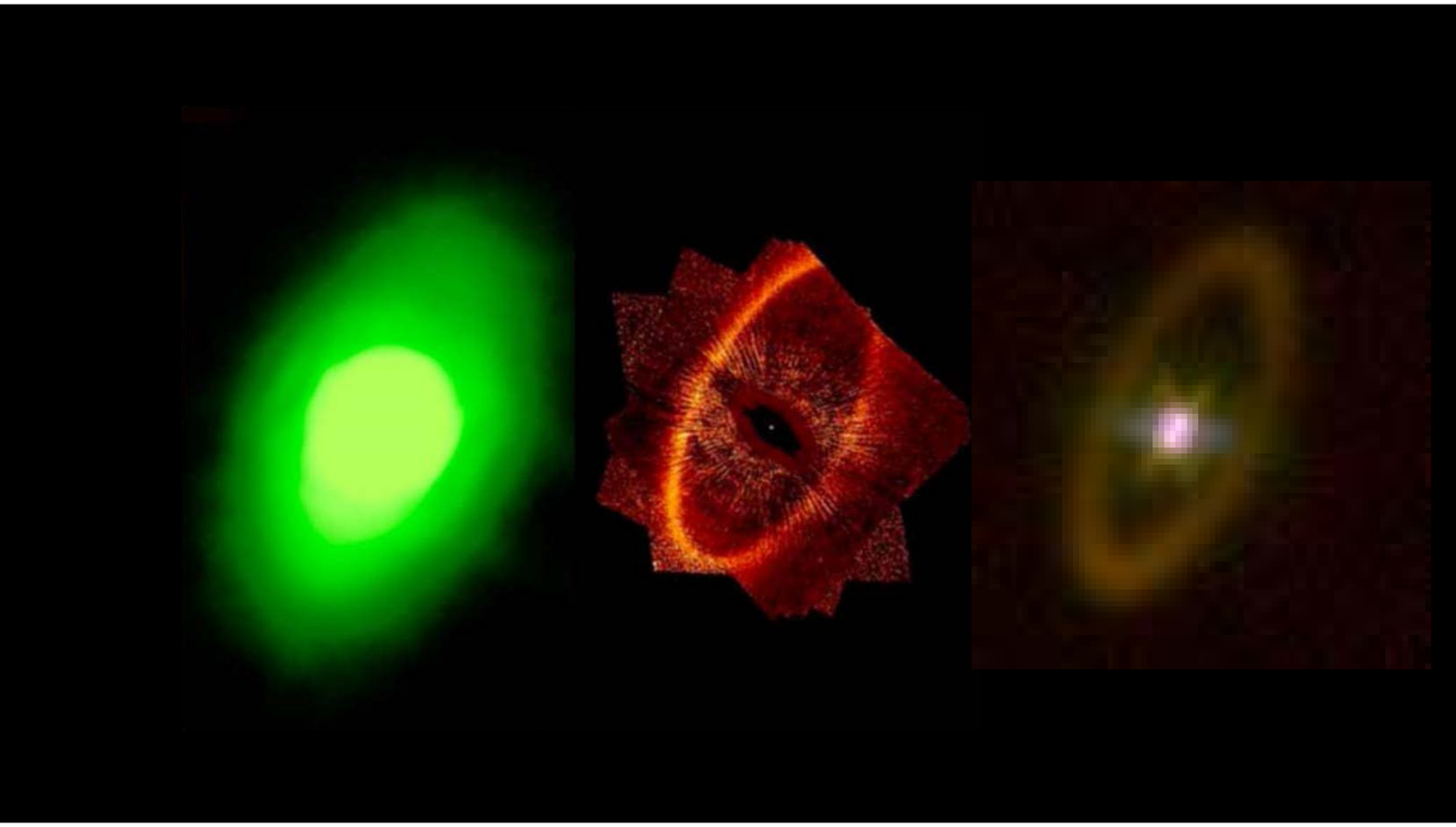}
\caption{LEFT: The emission for a 2 Myr year old 2 \mjup\ planet for different values of initial entropy (from the brighter ``hot" start to fainter ``cold" start) based on models by Spiegel \& Burrows (2012). The red rectangles across the top of the figure denote the location of selected NIRCam and MIRI coronagraphic filters. RIGHT:  A comparison of the Spitzer 24 \um\ and HST visible light images of the Fomalhaut disk with a simulation of a MIRI composite image at 15.5, 23 and 25.5$\mu$m (courtesy A. Gaspar). The disk seen at visible wavelengths is fully resolved with $\sim$1\arcsec\  resolution in the MIRI image.}
\label{fig:spectrum}  
\end{figure*}

The coronagraphs will be used to characterize previously imaged planets across a suite of NIRCam and MIRI filters and to discover new ones as well. From this information, it will be possible to infer a planet's  total luminosity, effective temperature, and thus effective radius, making it possible to estimate its  initial entropy  which is a direct indicator of its formation mechanism via a ``hot" or ``cold" start process (Figure~\ref{fig:spectrum}). JWST's sensitivity is such that it can detect known young planets at separations of $\leq$1\arcsec\ in just a few minutes of integration time.  NIRCam's medium passband filters will be used to characterize exoplanet atmospheres,   measuring both the continuum and searching for signatures of CH$_4$, CO and CO$_2$. MIRI's three 4QPM filters will isolate a band of NH$_3$ expected in cool ($<$1000 K) objects.

Observations using the  NIRSpec slit or IFS (R$\sim$ 1,000) and with the MIRI Low and Medium Resolution Spectrometers (LRS, R$\sim$100; MRS, R$\sim$2500)  will yield broad wavelength spectra  of  more widely separated planets. Some of these targets will be challenging due to the proximity of the host star, e.g. the  HR8799 planets. Others will be straightforward due to the faintness of the host star, e.g. 2MASS1207-3239, or to the very wide separation between the planet and host, e.g.  HD106906 (7\arcsec), GU Psc (42\arcsec) and WD0806-66 (130\arcsec). 

The Aperture Masking Interferometer capability within the  NIRISS instrument exploits very high angular resolution (0.5$\lambda/D$) to observe younger, brighter exoplanets located in more distant star forming regions. These include proto-planets -- possibly still accreting material -- in transition disks such as LkCa15 (Sallum et al 2015).

At separations outside $\sim$1\arcsec --  2\arcsec\, NIRCam's unmatched  sensitivity at 5 \um\  will reach much lower planet masses than is possible from the ground. Hour-long observations at 4.5 $\mu$m  can identify planets with masses below that of Saturn's mass in previously known planetary systems. Similar searches will be made for  $\sim$1 \mjup\ planets within 2--10\arcsec\ radius around  3 of the nearest, brightest debris disk systems. Furthermore, searches of the youngest and closest M stars (often too faint for ground-based adaptive optics imaging) offer the prospect of detecting planets with masses as low as a Uranus mass in orbits as small as 10-20 AU where such planets are thought to form (Schlieder  et al 2015). 

%\subsection{Debris Disk and Planets}

{\bf 3.2 Debris Disk and Planets:} JWST will open a dramatic new era in the study of debris disks and their interactions with planets. There is a rich literature of models describing the interaction between planets and disks  creating multiple rings,  gaps, and various asymmetries in our own and other planetary systems (e.g., Wyatt  2006,2008;   Ertel et al 2012). The planets responsible for these effects can range in size from  Uranus  to Jupiter  and even down to  terrestrial-sized for structures in asteroidal or exo-zodiacal disks (Stark et al 2008) with $\beta$ Pic (Lagrange et al 2010) as an example of a warped disk and HD202628 as an example of eccentric, shepherded ring system (Krist et al 2012). NIRCam and MIRI will play complementary roles in  advancing our understanding in this area. NIRCam will look for the planets while MIRI's coronagraphs will provide  new images of the closest debris disk systems.  Figure~\ref{fig:spectrum}  shows a sequence of images of the Fomalhaut disk as seen by Spitzer and HST along with a simulation of an image from JWST/MIRI. The offset of the ring  from its  star has been attributed to the interaction with one or more planets (Ertel et al 2012).
 
 %vskip -0.3in
\section{Transiting Planets}
 %vskip -0.15in
{\bf 4.1 JWST Potential for Transit Spectroscopy:} $JWST$ has  6.2 times the collecting area  of $HST$ and almost exactly 50 times the collecting area of Spitzer. This means that $R \simeq 200$ JWST spectra will have comparable S/N to $HST$ $R \simeq 35$ near-IR spectra and Spitzer IRAC $R \simeq 4.5$ mid-IR photometric observations of equal length. Thus JWST will produce considerably higher resolution and broader wavelength spectra than obtained by HST and will also provide the first high S/N mid-IR emission and transmission spectra of exoplanet atmospheres.

\begin{figure}[ht!]
\includegraphics[width=0.4\textwidth, angle=0]{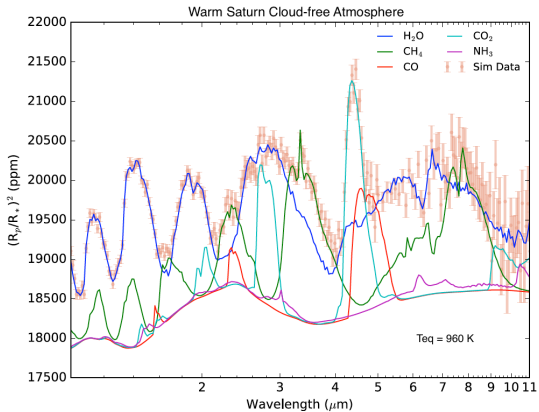}
 \caption{ \label{fig:chemistry_spec_fig} 
\small  Equilibrium molecular feature strengths and a simulated $JWST$ spectrum of a warm ($T_{\rm eq} = 960$ K) sub-Saturn mass planet with a H-dominated, cloud-free  atmosphere (HAT-P-12 b system parameters).  Model spectra were created using the CHIMERA suite (Line et al. 2013, 2014), and simulated JWST observations for 1 transit at each wavelength are shown. JWST spectra will probe the transition from CO  to CH$_4$ dominance in exoplanet atmospheres and will sensitively detect non-equilibrium chemistry (courtesy  E. Schlawin).
}
\end{figure}

JWST spectra of transiting planets will span numerous molecular features over wavelengths from less than 1 to more than 10 $\mu$m. From the strengths of the common molecules,  CH$_4$, CO, CO$_2$, H$_2$O, and NH$_3$ (not yet detected), one can infer a planet's C/O ratio and overall metal abundance, constraining its formation history (e.g., Brewer et al. 2017). JWST will be able to obtain high quality transmission and emission spectra of numerous warm-to-hot planets down to $\sim$ 10 Earth masses or less (Greene et al. 2016; Rocchetto et al. 2016; Molliere et al. 2017), advancing exoplanet characterization into a new era. Figure~\ref{fig:chemistry_spec_fig} shows 1.1 -- 1.7 $\mu$m  H$_2$O absorptions observable by HST as well as longer wavelength 1.7 -- 11 $\mu$m features of CH$_4$, CO, CO$_2$, and NH$_3$ that vary with temperature and will be diagnostic of the atmospheric composition and chemistry of cool to hot  (400 K $\leq T_{\rm eq} \leq$ 3000 K) planetary atmospheres.

JWST's instruments are well suited to transit spectroscopy: a) {\it NIRISS SOSS} mode uses a unique grism  to produce a slitless spectrum spanning 0.6 -- 2.8 $\mu$m in 2 orders simultaneously with  $R \equiv  \lambda / \delta \lambda \simeq 700$ resolving power; {\it NIRCam} has Si grisms that produce $R = 1100 - 1700$ slitless spectra over 2.4 -- 5.0 $\mu$m wavelengths.  We expect that the NIRCam dispersed Hartmann sensor (DHS) will be available for scientific use starting in Cycle 2. This will allow simultaneous 1 -- 2 $\mu$m and a subset of 2.4 -- 5.0 $\mu$m spectroscopy (Schlawin et al. 2017); {\it MIRI LRS} provides a slitless mode that acquires low resolution ($R \sim 100$) spectra over 5 -- $\sim12$ $\mu$m wavelengths (Kendrew et al. 2015; {\it NIRSpec} provides a wide, $1.6^{\prime\prime} \times 1.6^{\prime\prime}$ slit that is intended to minimize slit modulation of spectra in the presence of telescope jitter (Ferruit et al. 2014) and isolates targets from nearby stars. All of the 6 NIRSpec gratings and its 0.6 -- 5 $\mu$m  prism can be used for time series transit spectroscopy with this aperture.

%%%%%%%%%%%%% figure %%%%%%%%%%%%%%%%%%%%
%%%%%%%%%%%% figure %%%%%%%%%%%%%%%%%%%%

%\subsection{Archetypal Transiting Systems }
%vskip -0.15in
{\bf 4.2 Archetypal Transiting Systems:} Several studies have simulated $JWST$ spectra for a variety of planet types and have assessed what can be learned from these data from direct examination of spectra (Deming et al. 2009; Barstow \& Irwin 2016; Molliere et al. 2017; Morley et al. 2017) or statistical Bayesian retrieval techniques (Greene et al. 2016; Barstow et al. 2016; Batalha \& Line 2017; Howe et al. 2017). %We now draw from these works to discuss the expected data quality and molecular features that may be detectable in $JWST$ spectra of several different types of exoplanet atmospheres. The example spectra shown in this subsection are produced from the simulations by \citet{GLM16}, but \citet{MvBB17} and the other works cited above present numerous other examples.

\begin{figure*}
%\begin{tabular}{cc}
\includegraphics[width=0.5\textwidth]{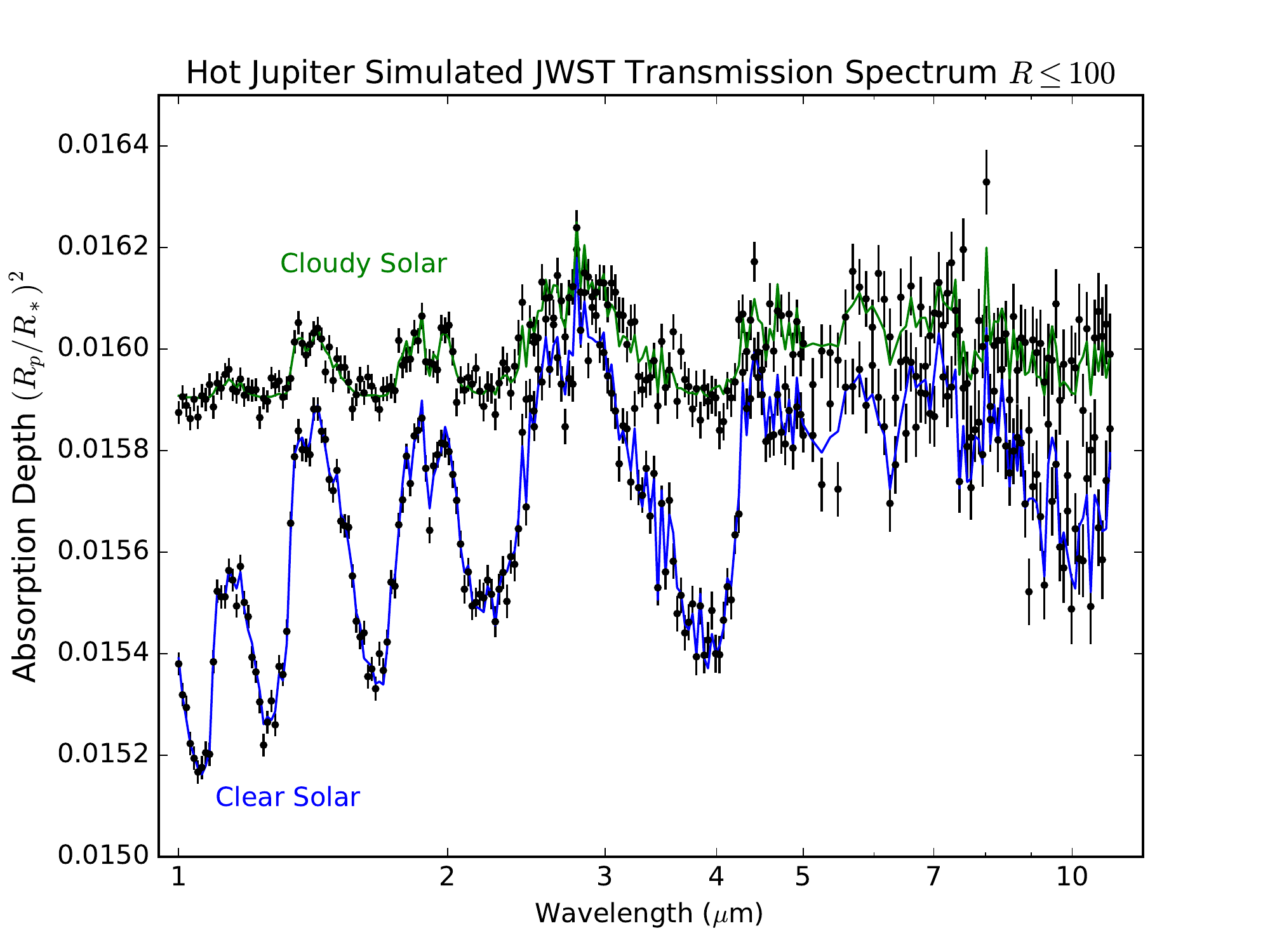}\includegraphics[width=0.5\textwidth]{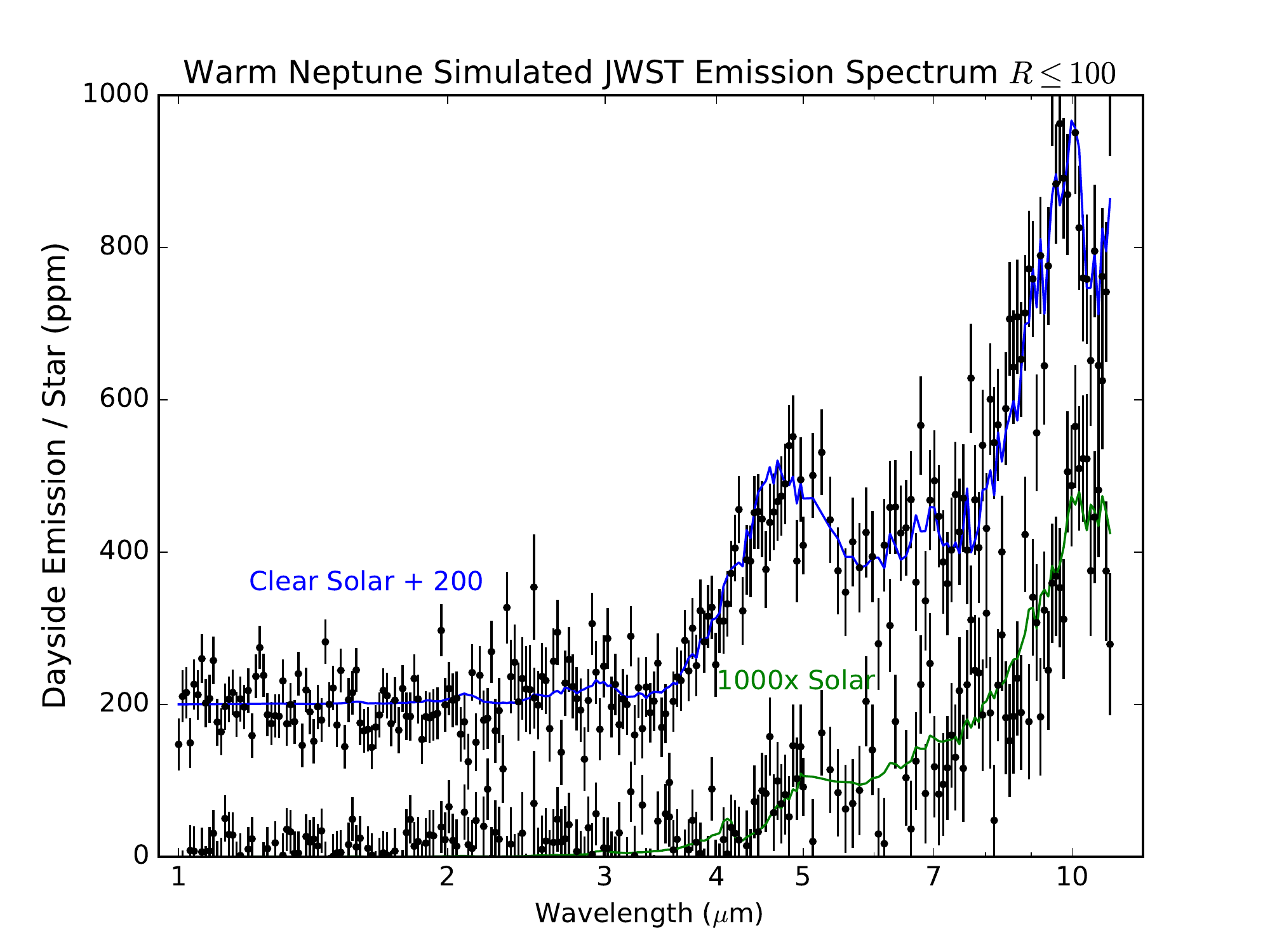}
%\plottwo{Hot_Jupiter_Solar_1transit_R100.pdf}{Warm_Neptune_1eclipse_R100.pdf}
%\includegraphics[width=1.0\textwidth, angle=0]{Hot_Jupiter_Solar_1transit_R100.pdf}
 %\end{tabluar}
 \caption{ \label{fig:hotJupiter} 
\small LEFT: Clear (blue curve) and cloudy (green curve) chemical equilibrium  models of a hot Jupiter planet with solar composition. HD 209458 b system parameters were used.
 RIGHT: Solar abundance (blue curve) and high mean molecular weight (1000$\times$ solar, green curve) emission spectra of a warm Neptune planet in chemical equilibrium. GJ 436 b system parameters were used.  BOTH: Simulated $JWST$ NIRISS SOSS, NIRCam grism, and MIRI LRS data are shown as black points and error bars and  binned to $R \leq 100$, with one transit or secondary eclipse simulated at each wavelength.
}
\end{figure*}

{\it Hot Jupiters} with bright host stars will generally have high S/N transmission and emission spectra in a single transit or secondary eclipse. Figure~\ref{fig:hotJupiter} shows that a hot Jupiter like HD 209458 b would have absorption features of several hundred parts per million (ppm) if its atmosphere was cloud-free, and these would be detected at high S/N. Even a cloudy atmosphere (opaque cloud at 1 mbar) would show $\sim100$ ppm features of H$_2$O, CO, and CO$_2$ that would be detected easily.

{\it Warm Neptune} planets with masses $M \sim 20M_{\oplus}$ and equilibrium temperatures 600 K$ < T_{\rm eq} <$ 1000 K will also be amenable to characterization with $JWST$ transmission and emission spectroscopy. Transmission feature strengths can also be several hundred ppm in size for M dwarf host stars, and strong H$_2$O and CH$_4$ features can be detected at high S/N for bright host stars, $K \leq \sim7$ mag. Figure~\ref{fig:hotJupiter} shows that a $T_{\rm eq} = 700$ K planet like GJ 436 b will   have a detectable emission spectrum at $\lambda > 4$ $\mu$m. H-dominated mini-Neptune planets ($R \simeq 2 R_\oplus$) with bright M star hosts are also expected to have similarly strong transmission and emission features.

{\it Super-Earth and Earth-sized} planetary atmospheres will be difficult but not impossible to characterize with $JWST$; numerous transmission or emission spectra will likely need to be co-added to make confident detections of their weak features (Morley et al. 2017).  Some TRAPPIST-1 planets  will be observed multiple times in transmission with NIRISS or NIRSpec or in emission with MIRI (at $\lambda > 10$ $\mu$m) in an effort to detect atmospheric features and study their thermal emission.

%%%%%%%%%%%%% figure %%%%%%%%%%%%%%%%%%%%
%%%%%%%%%%%% figure %%%%%%%%%%%%%%%%%%%%

%%vskip -0.15in
{\bf 4.3 Expected Characterization Results: } {\it Atmospheric compositions}  will be constrained with transmission spectra of warm Neptunes to hot Jupiters which  have clear atmospheres with solar to $\sim 100 \times$ solar metallicities (0 $\leq$ [M/H] $\leq$ 2).  Using information retrieval techniques, Greene et al. (2016) show that H$_2$O and CO (the dominant molecular components) can be constrained to factors of 1.5 -- 2 with transmission spectra of hot Jupiter planets in favorable systems. The mixing ratios of H$_2$O and CH$_4$, the dominant molecules in cooler atmospheres, can be constrained to a factor of 3 -- 10 for warm Neptune and sub-Neptune planets with clear atmospheres. %These results are for a single transit at each wavelength. Complete $<$1 -- 11 $\mu$m spectra provide the tighter constraints in these ranges, while NIRISS SOSS ($\lambda \leq 2.5$ $\mu$m) alone provides the looser ones. Emission spectra of these systems provide similar constraints for Hot Jupiters if the entire   $<$1 -- 11 $\mu$m spectrum is observed, but this degrades to a factor of $\sim4$ if only NIRISS SOSS spectra are acquired. Emission spectra of warm Neptunes and sub-Neptunes can constrain $H_2$O and CH$_4$ mixing ratios to only a factor of $\sim10$. 

{\it C/O ratios and metallicity} provide clues to a planet's formation history.  In addition to C/O (e.g., Brewer et al. 2017), overall planet metal enrichment [M/H] may also be related to formation scenarios. Gas and ice giant solar system planets have [M/H] values that increase by a factor of $\sim30$ as their masses decrease from Jupiter to Uranus (Kreidberg et al. 2014). Planet formation scenarios can be investigated by measuring the C/O ratios of exoplanet atmospheres with JWST. Greene et al. (2016) show that $JWST$ spectra will likely be able to constrain C/O ratios to within $\sim50$\% for transmission or emission spectra of clear to cloudy hot Jupiters obtained over a single transit or secondary eclipse at all wavelengths 1 -- 11 $\mu$m. %Warm Neptunes and sub-Neptunes are somewhat worse, with typical 1 $\sigma$ C/O constraints from $\sim50$\% to a factor of $\sim3$ in most transmission scenarios. They also find that full-wavelength $JWST$ spectra constrain [M/H] values to about a factor of 3 in these planets, so the enrichment seen in the solar system should be detected in $JWST$ spectra with high confidence.

{\it 3-D thermal profiles, chemical disequilibrium, and cloud properties} will also be revealed in $JWST$ spectra. Emission spectra will reveal dayside temperature-pressure profiles over a large pressure range $\sim1$ bar (e.g., Line et al. 2014), and transmission spectra will measure the temperatures and compositions in planet terminators. Phase curve spectra will also reveal global circulation patterns in these parameters. Chemical disequilibrium due to vertical mixing or photochemistry (e.g., Line \& Yung 2013) should also be detectable. Finally, cloud particles will contribute to the mid-IR emission spectra of hot planets (e.g., Crossfield 2015), and cloud-top pressures can be retrieved from transmission spectra (e.g., Greene et al. 2016).

{\bf 4.4 Characterizing Terrestrial Atmospheres:}  M dwarf stars will be the most favorable hosts for characterizing Earth-sized planets  via transit techniques. Transmission spectroscopy of true Earth analogs (size, temperature, composition) will be exceptionally difficult because of the minuscule signals ($\sim$10 ppm or less; Schwieterman et al. 2016). However, transmission signals are expected to be larger and potentially detectable for several known warmer Earth-sized planets if they have non-Earth-like atmospheric compositions (Morley et al. 2017). The thermal emission of the $T_{\rm eq} = 400$ K TRAPPIST-1 b planet is expected to be about 300 ppm relative  to its star at wavelengths 10 -- 15 $\mu$m.  Planned $JWST$ MIRI observations are expected to detect this emission at S/N $\geq$ 5 in 5 secondary eclipses in the 12.8 $\mu$m and 15 $\mu$m filters. Detection at 12.8 $\mu$m and non-detection at 15 $\mu$m would indicate a strong Earth-like atmospheric CO$_2$ feature. In principle, dozens of observations could be co-added to search for the $\sim 50$ ppm mid-IR emission signals expected for $T_{\rm eq} < 300$ K planets (e.g., TRAPPIST-1 d and e). However this requires significantly better performance than JWST requirements and must be verified  on-orbit. 

 %vskip -0.3in
\section{Conclusion}
 %vskip -0.1in
JWST offers dramatic new spectroscopic and coronagraphic capabilities  for the study of planets ranging in mass from Earths to Jupiters in orbits from a few days to thousands of years. JWST observations will shift exoplanet science from its census-taking phase into a period of detailed physical characterization. Future missions like LUVOIR, HabEx and OST will someday obtain spectra of Habitable Zone Earth biomarkers, but JWST is essential now to  learn about the formation and evolution of planetary systems and the physical properties of individual planets.
\vskip 0.2in
Some of the research described in this publication was carried out in part at the Jet Propulsion Laboratory, California Institute of Technology, under a contract with the National Aeronautics and Space Administration.
Copyright 2018 California Inst of Technology. All rights reserved. 
 %vskip -0.3in
\section{References}
 %vskip -0.1in
\scriptsize{%\scriptsize
%Acke, B. et al. 2012, A\&A,  540, A125   $\Diamond$   
%Barstow et al 2015, MNRAS, 448, 2546 $\Diamond$  
Barstow et al 2016, MNRAS, 458, 2657 $\Diamond$  
Barstow, J. K., et al  2016, MNRAS,  461, L92 $\Diamond$  
Batalha, N. E. \& Line, M. R. 2017, AJ, 153, 151 $\Diamond$  
Beichman, C.A. et al.  2010, PASP, 122, 162 $\Diamond$  
Beichman, C.A et al.  2014, PASP, 126, 1134 $\Diamond$ 
Brewer, J. M. et al. 2017. ApJ, 153, 83  $\Diamond$  
Crossfield, I. PASP, 127, 941 $\Diamond$  Deming, D. et al. 2009,  PASP, 121, 952 $\Diamond$ 
Ertel, S. et al. 2012, A\&A,  544, A61 $\Diamond$  
Ferruit, P.  et al. 2014,  Proc. SPIE,   9143,  91430A $\Diamond$  
Gillon, M. et al. 2017, Nature, 542, 456 $\Diamond$  
Glasse, A.  et al. 2015,  PASP, 127, 686 $\Diamond$ Greene, T. et al.  2016, ApJ, 817, 17 $\Diamond$  
Howe, A. R., et al. 2017, ApJ, 835, 96 $\Diamond$  Kendrew, S., et al.  2015, PASP, 127, 623 $\Diamond$  Kreidberg, L. et al.  2014, ApJ, 793, L27 $\Diamond$  Krist, J. E., et al. 2012, AJ, 144, 45 $\Diamond$  Lagrange, A.-M.  et al. 2010, Science, 329, 57 $\Diamond$  Line, M. R.  et al,  2014, ApJ, 783, 70 $\Diamond$  Line, M. R., \& Yung, Y. L. 2013, ApJ, 779, 3 $\Diamond$  Line, M. R. et al. 2013,  ApJ, 775, 137 $\Diamond$ Marley, M. S.  et al. 2007, ApJ, 655,  541 $\Diamond$ 
Molliere, P. et al.   2017, A\&A, 600, A10 $\Diamond$  
Mordasini, C.   et al.  2016, ApJ, 832, 41 $\Diamond$  
Morley, C. V. et al. 2017, ApJ, 850, 121 $\Diamond$ 
Oberg, K. I., et al    2011, ApJ, 743, L16 $\Diamond$  
Rocchetto, M.   et al  2016, ApJ, 833, 120 $\Diamond$  
Sallum, S. et al.   2015, Nature, 527, 342 $\Diamond$  
Schlawin, E., et al. 2017, PASP, 129, 015001 $\Diamond$ 
Schlieder, J. E., et al.  2015, arXiv:1512.00053 $\Diamond$ 
Schwieterman, E. W. et al. 2016, ApJ, 819, L13 $\Diamond$  
%Sing, D. K. et al. 2016, Nature, 529, 59 $\Diamond$  
Spiegel, D. S. et al. 2012, ApJ, 745, 174 $\Diamond$  
Stark, C. C. et al. 2008, ApJ, 686, 637 $\Diamond$  
Wyatt, M. C. 2006, ApJ, 639, 1153 $\Diamond$  Wyatt, M. C.  2008, ARAA, 46, 339}.
%\bibliography{references}

\end{document}